\documentclass[5p,times,twocolumn,a4paper,showpacs,showkeys,amsmath,amssymb,superscriptaddress,nofootinbib]{elsarticle} 

\usepackage{lineno,hyperref}
\usepackage{comment}
\usepackage{upgreek}
\usepackage{amsmath}

\usepackage{siunitx}
\biboptions{sort&compress}

\usepackage[whole]{bxcjkjatype} 
\usepackage{color} 

\modulolinenumbers[5]

\newcommand{\figref}[1]{Fig.~\ref{#1}}
\newcommand{\tabref}[1]{Table~\ref{#1}}
\renewcommand{\eqref}[1]{Eq.~\ref{#1}}
\newcommand{\secref}[1]{Sec.~\ref{#1}}
\newcommand{\citeref}[1]{Ref.~\cite{#1}}

\newcommand{\meg}{\mu\to e\gamma}
\newcommand{\D}{\mathrm{d}}

\journal{Journal of \LaTeX\ Templates}

\begin{document}


\title{Transformer-Based Approach to Enhance Positron Tracking Performance in MEG~II}

\author[pisa]{Lapo Dispoto}
\author[liverpool]{Fedor Ignatov}
\author[icepp]{Atsushi Oya\corref{mycorrespondingauthor}}
\cortext[mycorrespondingauthor]{Corresponding author}
\ead{atsushi@icepp.s.u-tokyo.ac.jp}
\author[kek]{Yusuke Uchiyama}
\author[pisa]{Antoine Venturini}

\address[icepp]{International Center for Elementary Particle Physics, The University of Tokyo, 7-3-1 Hongo, Bunkyo-ku, Tokyo 113-0033, Japan}
\address[kek]{KEK, High Energy Accelerator Research Organization, 1-1 Oho, Tsukuba, Ibaraki 305-0801, Japan}
\address[liverpool]{Oliver Lodge Laboratory, University of Liverpool, Liverpool, L69 7ZE, United Kingdom}
\address[pisa]{INFN Sezione di Pisa$^{a}$; Dipartimento di Fisica$^{b}$ dell'Universit\`a, Largo B.~Pontecorvo~3, 56127 Pisa, Italy}

\begin{frontmatter}

\begin{abstract}
We developed a Transformer-based method for positron track reconstruction in the MEG~II experiment.
The model acts as a hit classifier to remove pileup hits in the MEG~II drift chamber, which operates under a high pileup occupancy of \SIrange{35}{50}{\percent}. 
The trained model significantly improved hit purity, leading to enhancements in tracking efficiency and resolution by \SI{15}{\percent} and \SI{5}{\percent}, respectively, at a muon stopping rate of \SI{5e7}{\mu/s}. 
This improvement translates into an approximately \SI{10}{\percent} increase in the sensitivity of the $\meg$ branching ratio measurement.
\end{abstract}


\end{frontmatter}


\section{Introduction}

\begin{figure}[tbp]
   \centering
   \includegraphics[width=\linewidth]{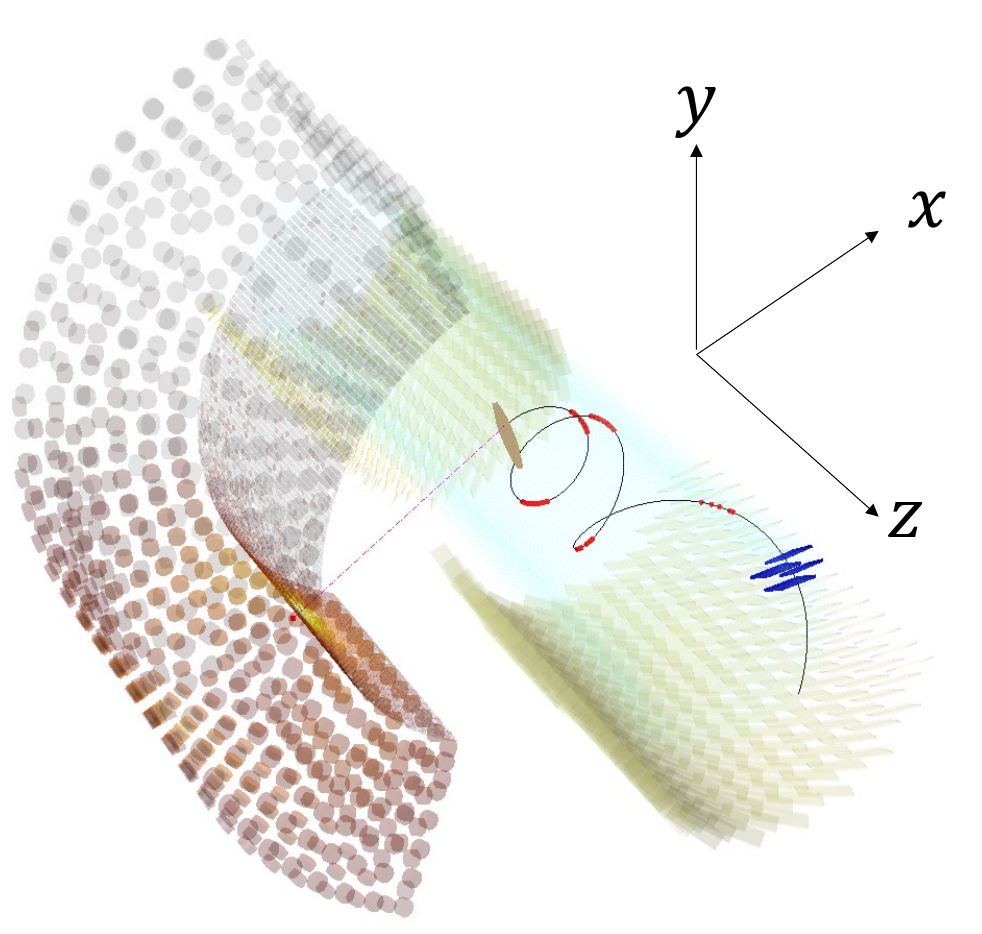}
   \caption{An event display for a simulated $\meg$ event, showing the coordinate definition together. 
            Inside the CDCH volume (cyan volume), the positron track (black solid line) makes 2.5 turns leaving red-marked hits. 
            After leaving the CDCH volume, the positron makes hits on four (blue solid tiles) out of the 512 scintillation counters of the pTC (yellow transparent tiles).
            The photon, emitted in the direction opposite to that of the positron, is also shown as the magenta dashed line, together with the C-shaped liquid xenon detector.
   }
   \label{fig:DisplayDetector}
\end{figure}

\noindent The MEG~II experiment searches for the charged lepton flavor-violating muon decay, $\meg$ \cite{MEGIIDetectorPaper2023}.
Data collection began in 2021, and analysis of the first two years of data (up to 2022) set a \SI{90}{\percent} C.L. upper limit of \num{1.5e-13} on the $\meg$ branching ratio \cite{MEGIIResult2025}.
The experiment is scheduled to continue until 2026, aiming to achieve an upper-limit sensitivity of \num{6e-14}, with its scientific motivation and impact discussed in \citeref{MEGIIDesign2018}.
Located at the $\pi E5$ beam line of PSI \cite{PiE5}, the experiment stops a continuous muon beam in a thin plastic target at a rate of \SIrange{3}{5e7}{\mu/s}, and measures positrons and photons from muon decays using a positron spectrometer and a liquid xenon detector, respectively (\figref{fig:DisplayDetector}).

In this paper, we aim to improve the performance of the positron spectrometer, thereby enhancing the $\meg$ sensitivity of the experiment.
The positron spectrometer must achieve high positron detection efficiency and resolution even at a high muon stopping rate on the target.
The efficiency and the stopping rate directly determine the number of muon decays measured effectively, while the resolution helps suppress background events.
The dominant source of background is accidental coincidences of positrons and photons from different muon decays, where the positron–photon pairs have neither time nor angular correlations and exhibit continuous energy spectra up to \SI{52.8}{MeV}.
With excellent resolution, such events can be kinematically distinguished from the two-body kinematics of the $\meg$ signal: a time-coincident positron and photon, each with \SI{52.8}{MeV} energy, emitted back-to-back.

The latest MEG~II result \cite{MEGIIResult2025} adopted the positron spectrometer analysis method described in \citeref{MEGIICDCHPaper2023}.
In that study, \citeref{MEGIICDCHPaper2023} concluded that a muon stopping rate of \SI{4e7}{\mu/s} maximizes the $\meg$ sensitivity.
This value is lower than the rate limit of \SI{5e7}{\mu/s}, which is determined by constraints from the liquid xenon photon detector.
The reduced muon rate was chosen because an increase in the stopping rate was observed to degrade the positron tracking efficiency.
Since the single-hit efficiency of the tracking detector remains nearly constant across different stopping rates, this degradation was attributed to an inefficiency in the reconstruction algorithm.
In particular, the tracking algorithm is inefficient in the track-finding stage, where pattern recognition is highly sensitive to contamination from pileup hits.
This underscores the importance of improving the algorithm to make it more robust against pileup.

This paper presents a novel Transformer-based machine learning method \cite{AttentionAllYouNeed} for tracker hit filtering, which enhances the performance of the MEG~II positron track reconstruction.
\secref{sec:Spectrometer} provides a brief introduction to the positron spectrometer design and the conventional reconstruction method detailed in \citeref{MEGIICDCHPaper2023}.
Next, \secref{sec:RelatedWork} reviews related work in the literature on applying machine learning techniques to track reconstruction and discusses the specific tracking requirements of MEG~II.
\secref{sec:TransformerTracking} describes the Transformer model architecture adapted for the MEG~II experiment.
\secref{sec:Performance} evaluates the tracking performance achieved with the proposed method.
Finally, \secref{sec:Discussion} discusses the results, and \secref{sec:Conclusion} summarizes the conclusions of this paper.

\section{Review of MEG~II positron spectrometer}\label{sec:Spectrometer}
\noindent The design, analysis methods, and performance of the positron spectrometer are detailed in \citeref{MEGIIDetectorPaper2023,MEGIICDCHPaper2023}.
In the following subsections, \secref{sec:Coordinate} introduces the definition of the coordinate system, which will be used throughout this paper.
The spectrometer hardware and the conventional reconstruction method are concisely described in \secref{sec:SpectrometerHardware} and \secref{sec:ConventionalReco}, respectively.
Finally, \secref{sec:ConventionalPerformance} explains the performance evaluation method.

\subsection{Notation and coordinate system}\label{sec:Coordinate}
\noindent The coordinate system, shown in \figref{fig:DisplayDetector}, is defined with its origin at the center of the MEG~II detector, where the muon stopping target is installed (See also Fig.~1 of \citeref{MEGIIDetectorPaper2023}).
The $z$ axis is aligned with the muon beam direction, and the $y$ axis points upward.
The remaining $x$ axis completes the right-handed coordinate system and is oriented opposite to the C-shaped liquid xenon photon detector.
In this paper, $r$ denotes the radial coordinate in the usual cylindrical system, and $(\theta, \phi)$ denotes the coordinates in the spherical system.

\subsection{Spectrometer hardware}\label{sec:SpectrometerHardware}

\noindent The MEG~II positron spectrometer consists of a cylindrical drift chamber (CDCH) for tracking and a pixelated scintillation timing counter (pTC) for timing measurements, both placed inside a gradient magnetic field generated by the COBRA superconducting magnet.
As exemplified by the black line in \figref{fig:DisplayDetector}, the helical positron trajectories within the magnet include multiple turns along their flight from the muon stopping target to the pTC, and these must be reconstructed without missing any segment.
Hereafter, a \emph{single turn} refers to a pair of track segments: one outgoing (i.e., increasing in $r$) and one incoming, whereas a \emph{half turn} refers to either the incoming or outgoing segment alone.
About \SI{85}{\percent} of positrons complete 1.5 turns before reaching the pTC, while the remaining tracks, with a positron emission angle of $\theta_e \sim \SI{90}{\degree}$, make 2.5 turns or, more rarely, 3.5 turns.

The COBRA magnet generates a gradient magnetic field ranging from \SI{1.27}{T} at the center to \SI{0.5}{T} at the ends.
This gradient field configuration enhances (or suppresses) the track radius $r$ of positrons emitted with small (large) $\sin\theta_e$ and correspondingly small (large) transverse momenta.
As a result, the track radii of positrons with the same energy remain nearly constant regardless of the emission angle, which helps in selecting positrons close to \SI{52.8}{MeV}.

The CDCH (the cyan volume in \figref{fig:DisplayDetector}) is a single-volume wire chamber with an inner radius of $r = \SI{17}{cm}$, an outer radius of $r = \SI{29}{cm}$, and a length of \SI{191}{cm}.
It consists of nine drift-cell layers arranged in a stereo configuration, where the wire direction alternates between neighboring layers, enabling precise $z$ determination during track reconstruction.
Each layer contains 192 cells arranged symmetrically in $\phi$ rotation.
For each hit, the track impact parameter relative to the wire is determined from the drift time with a resolution of \SI{150}{\micro m}.
The maximum drift time is approximately \SI{300}{ns}, within which pileup hit occupancy in each cell reaches \SIrange{35}{50}{\percent} (lower occupancy in the outer layers) when operating at a muon stopping rate of \SI{5e7}{\mu/s}.
In addition, the hit position along the wire direction (corresponding to $z$ when neglecting the stereo angle) is measured with a resolution of \SI{7.5}{cm} using the charge ratio and time difference of signals detected at both ends.
Although the track $z$ parameter is primarily determined by the geometrical constraints of the stereo configuration, the \SI{7.5}{cm} hit resolution helps mitigate the effects of high pileup occupancy.

The pTC consists of 512 plastic scintillation counters (the yellow transparent tiles in \figref{fig:DisplayDetector}), each coupled to silicon photomultipliers (SiPM) on both ends.
These counters are divided into upstream and downstream sectors, which are mirror-symmetric to each other.
Each sector is pixelated along the $\phi$ and $z$ directions, forming a $16 \times 16$ array arranged symmetrically in $\phi$ rotation.
The positron impact time on each counter can be measured with an average precision of \SI{110}{ps}, and the positron emission time from the muon stopping target is reconstructed by combining multiple hits, each corrected for time-of-flight (TOF).
For signal positrons with energy \SI{52.8}{MeV}, the average number of pTC hits is 9, resulting in an overall positron time resolution of about \SI{40}{ps}.
An example hit cluster is shown as the blue solid tiles in \figref{fig:DisplayDetector}.

\subsection{Conventional positron reconstruction}\label{sec:ConventionalReco}

\noindent Positron reconstruction begins with the analysis of pTC hits.
Hits are clustered into groups based on their timing and position, which separates pileup hits originating from different positrons (See Sec.~5 of \citeref{MEGIIDetectorPaper2023}).
Thanks to the time resolution of each hit and the segmented detector design, this clustering algorithm is robust against pileup hits, with the inefficiency from mis-clustering being below \SI{1}{\percent}.
After clustering, the hit position within each counter region is estimated with a precision of approximately \SI{1}{cm}, using the time difference between the two SiPM readout channels and the hit pattern within the cluster.
The cluster time and position are then used in the subsequent tracking algorithm with the CDCH.

The first algorithm using CDCH hits performs pattern recognition for track finding, aiming to identify sets of hits belonging to the same turn segment (either a full single turn or a half turn) of positron trajectories.
The initial step constructs a track seed from four hits, with two neighboring layers each providing two nearby hits.
A valid seed corresponds to a trajectory that consistently resolves the left–right ambiguity and satisfies the drift distance for all four hits.
For drift distance determination, the pTC cluster provides a time reference for calculating the drift time, with a \SI{5}{ns} uncertainty due to TOF.
Once seeds are found, they are propagated forward and backward through the CDCH layers, adding new hits when they are consistent with the trajectory.
Propagation continues until reaching the boundary of the active volume, thereby forming a single- or half-turn candidate.
For clarity in later sections, note that this method does not exploit patterns between distant hits, particularly those belonging to different turn segments.

The candidate track segments identified by pattern recognition are fitted and then merged with other turn segments to reconstruct multi-turn trajectories.
The fitting algorithm uses the deterministic annealing filter implemented in the GENFIT package \cite{GenFitRef1,GenFitRef2}.
During the merging step, trajectories are propagated forward and backward using the refined track fit results, and turn segments with consistent kinematics are combined.
Further tracking refinements are then applied, including updating the drift distance with a TOF-corrected reference time, recovering missing hits in the fitted tracks, matching CDCH tracks with pTC clusters, and performing track selection.
Although this refinement step is essential for tracking performance, it is less relevant to this paper, which focuses primarily on improving the efficiency at the pattern recognition stage. 
Readers can refer to \citeref{MEGIICDCHPaper2023} for details and to \citeref{MEGIIResult2025} for recent updates.

\subsection{Tracking performance evaluation}\label{sec:ConventionalPerformance}
\noindent The tracking efficiency is evaluated by counting the number of reconstructed and selected tracks in a dedicated dataset collected with an unbiased trigger logic. This data-driven method is also used to normalize the dataset when calculating the $\meg$ branching ratio in \citeref{MEGIIResult2025}.
The efficiency is computed by dividing this count by the expected number of positron tracks emitted in the direction opposite to the acceptance region of the liquid xenon photon detector. This expectation is obtained as the product of the muon stopping rate on the target, the branching ratio of positrons in the energy range of interest, and the detector acceptance.
In \citeref{MEGIICDCHPaper2023}, using the tracking method described in \secref{sec:ConventionalReco}, the tracking efficiency was reported as $\SI{66\pm 4}{\percent}$ ($\SI{77\pm 4}{\percent}$) at a muon stopping rate of \SI{5e7}{\mu/s} (\SI{2e7}{\mu/s}), indicating efficiency degradation at higher pileup rates.
The quoted uncertainty is dominated by that of the muon stopping rate (see \citeref{MEGIIDetectorPaper2023} for its evaluation), which enters the denominator of the efficiency calculation.

The evaluation of tracking resolution uses two methods: fitting the Michel end-point spectrum of the muon Standard Model decay $\mu \to e \nu \nu$, and \emph{double-turn analysis}.
The first method evaluates the momentum resolution and is fully data-driven.
The fit compares the observed Michel end-point spectrum in data with the sharp \SI{52.8}{MeV} upper edge of the theoretical Michel spectrum \cite{Kinoshita:1958ru}, which in practice is smeared by the finite detector resolution.
The fit function is obtained by multiplying the theoretical spectrum by an efficiency function, which accounts for variations in detector acceptance at different momenta, and then convolving it with a resolution function.
The second method, double-turn analysis, evaluates the resolution of angle and position at the positron emission point on the muon stopping target.
Tracks with 2.5 turns are split into the first single turn and the last 1.5 turns, and their kinematics are compared after fitting each turn individually.
This method requires corrections for differences in the number of hits between the original and split tracks, which involves MC-based adjustments.
The most up-to-date resolution estimates are reported in \citeref{MEGIIResult2025}.

\section{Related machine learning works for track reconstruction}\label{sec:RelatedWork}

\subsection{Machine learning technique for charged particle tracking} 
\noindent Machine learning techniques for charged particle tracking have been extensively studied in the literature and are summarized in the living review \cite{MLLivingReview}.
Among these, graph neural networks (GNNs) are the most widely adopted approach
\cite{Correia:2024ogc,ExaTrkX:2021abe,Reuter:2024kja,Caillou:2024smf,Caillou:2024dcn,ExaTrkX:2020apx,Liu:2023siw,Lieret:2023aqg,Lieret:2023ydc,Akram:2022zmj,Jia:2024rbx}, due to their ability to model irregular detector geometries and exploit relational information between hits.
Although these models have demonstrated promising performance in various experiments, they suffer from a critical bottleneck in the initial graph construction stage, which becomes increasingly expensive as pileup occupancy rises.
More recently, Transformer-based models --- motivated by their success in computer vision and natural language processing and inspiring the approach in this study --- have emerged for vertexing and tracking tasks \cite{VanStroud:2024fau,VanStroud:2023ggs,Huang:2024voo,Melkani:2024yuj}.
These approaches aim to overcome the combinatorial complexity of tracking by leveraging attention mechanisms.

\subsection{Specific demands of the MEG~II tracking task} 
\noindent This work aims to develop an ML-based hit filtering method for the MEG~II CDCH to overcome tracking performance limitations caused by the harsh pileup environment, which restricted the muon stopping rate on the target to \SI{4e7}{\mu/s} \cite{MEGIICDCHPaper2023}.
However, applying ML techniques developed in other studies is not straightforward for the MEG~II drift chamber.
Unlike silicon detectors used at the LHC, for example, the drift chamber cannot directly measure the three-dimensional position of hits, making tracking more challenging as it relies only on wire position and drift distance estimation.
Even compared with other experiments using drift chambers, the MEG~II chamber is characterized by a high cell occupancy of \SIrange{35}{50}{\percent} within the drift time window.
In addition, positrons in MEG~II traverse only up to nine drift-cell layers, resulting in shorter path lengths that can easily be obscured by pileup hits.
Moreover, TOF measurement requires complete tracking of multiple-turn trajectories between the muon stopping target and the pTC, even though these trajectories exit the CDCH volume ($17 < r < \SI{29}{cm}$) between different turn segments and leave no hits there.
As a comparison, \citeref{Reuter:2024kja} discusses tracking a similar topology, referred to as ``curling tracks'', where the trajectories found by the GNN are not associated with hits after re-entering the chamber volume.

\section{Positron tracking with Transformer model}\label{sec:TransformerTracking}

\subsection{Concept of the model design}\label{sec:TransformerModelConcept}
\noindent In this study, we developed a Transformer-based model for the CDCH hit filtering task, which leverages global patterns of pTC and CDCH hits across different turn segments.
The key idea is that an algorithm exploiting correlations between distant hits is more robust against pileup than the local algorithm described in \secref{sec:ConventionalReco} (i.e., connecting nearby CDCH hits on the same turn).
By taking advantage of the separation of pileup pTC hits through the clustering algorithm, the model aims to identify CDCH hits belonging to the positron track associated with a given pTC cluster.
One inference is performed for each pTC cluster, and this process can be repeated multiple times within an event.

\begin{figure}[tbp]
   \centering
   \includegraphics[width=\linewidth]{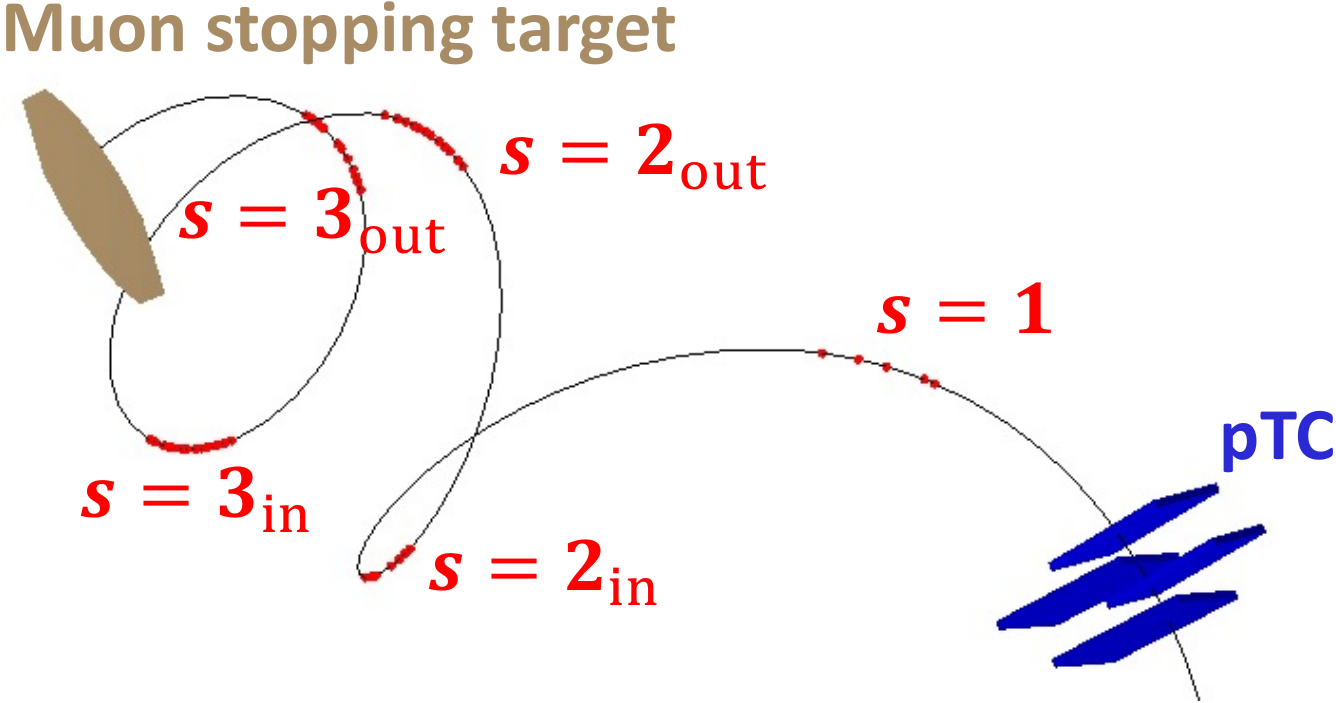}
   \caption{CDCH hit label for the 2.5 turn positron trajectory presented in \figref{fig:DisplayDetector}. 
            Red markers, indicating CDCH hits, are labelled by the turn segment it belongs to.
            Emitted from the muon stopping target (brown), the positron leaves $s=3$ hits on the first turn, then $s=2$ hits on the second turn, and finally $s=1$ hits on the last half turn before leaving hits on the pTC (blue tiles).} 
   \label{fig:DisplayTrajectory}
\end{figure}

The model is designed as a classifier with multiple output class labels, $s \in \{\varnothing, 1, 2_\mathrm{in}, 2_\mathrm{out}, 3_\mathrm{in}, 3_\mathrm{out}, 4_\mathrm{in}, 4_\mathrm{out}\}$.
The $\varnothing$ label corresponds to pileup or noise hits, while the other labels represent hits on the track of interest, defined by the track turn segment as shown in \figref{fig:DisplayTrajectory}.
Here, the $s$ index increases as the track is back-propagated by one turn: $s = 1$ corresponds to the last half-turn before the pTC, $s = 2$ to the preceding turn, and so on.
The subscripts \emph{in} and \emph{out} divide a single turn into its incoming and outgoing half-turn segments.
With the maximum $s$ set to four, the model targets tracks of up to 3.5 turns, as $\meg$ events with positron tracks exceeding 3.5 turns would not be efficiently collected by the time-coincidence trigger logic (see \citeref{MEGIIDetectorPaper2023}).
Note that the model only distinguishes signal hits from pileup hits --- it neither sorts CDCH hits along the track direction nor directly estimates track kinematics.

The model output is then used as a hit filter to discard pileup CDCH hits; only the selected hits are used for track candidate construction and fitting, which reuses the methods described in \secref{sec:ConventionalReco}.
The hit selection and track candidate construction procedure is repeated turn-by-turn; one attempt is made for hits predicted to be $s=1$, another for those with $s=2\cdots$.
This is because hits belonging to different segments contaminate the track candidate construction algorithm, which can associate only hits on the same turn.
The hit selection criteria for the $s=i$ track turn segment are
\begin{equation}\label{eq:HitFilteringCriteria}
  \begin{cases}
    f_\mathrm{hit}(s=1) > f_\mathrm{th} & (i = 1),\\
    f_\mathrm{hit}(s=i_\mathrm{in}) > f_\mathrm{th} \lor f_\mathrm{hit}(s=i_\mathrm{out}) > f_\mathrm{th} & (i \in \{2,3,4\})
  \end{cases}
\end{equation}
where $f_\mathrm{hit}(s)$ is the $s$-prediction score of each CDCH hit, and $f_\mathrm{th}$ is a threshold defined later in \secref{sec:EfficiencyPurity}.
In addition to the use in the track finding, the selected hits and their labels are referred to again during the tracking refinement after the initial track fitting, attempting to recover hits that are selected by the model but not included in the fitted tracks.

\subsection{Model architecture}
\begin{figure}[tbp]
   \centering
   \includegraphics[width=\linewidth]{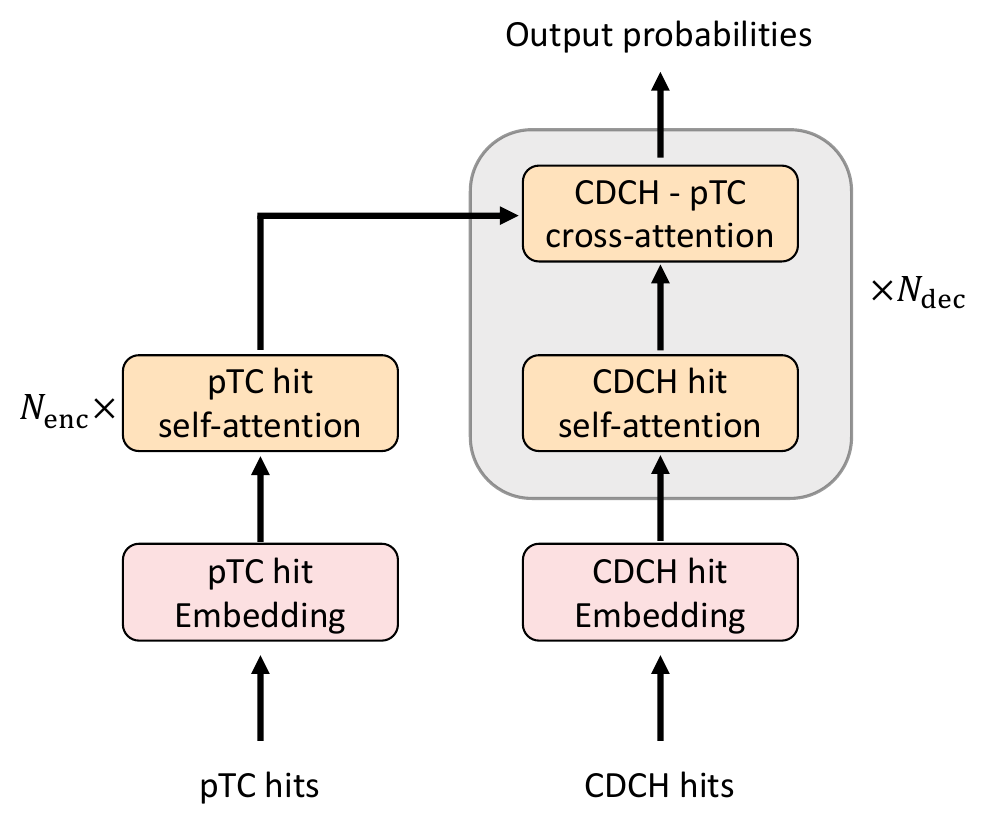}
   \caption{Transformer model adapted to the MEG~II positron spectrometer. Two sets of queries, pTC hits and CDCH hits, are input to the model. The model outputs the probability of each CDCH hit being associated with the pTC hits.}
   \label{fig:ModelArchitecture}
\end{figure}

\noindent The architecture of the Transformer model is shown in \figref{fig:ModelArchitecture}.
The attention blocks are divided into two parts: one for pTC and one for CDCH, with model outputs attached only to the latter.
The pTC block processes hits belonging to a specific pTC cluster, and its outputs cross-attend to CDCH hits.
The CDCH block receives all CDCH hits in an event, performs self-attention among CDCH hits followed by cross-attention with pTC hits, and finally outputs the class probabilities.
The key idea is that connections between CDCH hits are identified through self-attention, including those from different turn segments, while CDCH hits are matched to pTC hits via cross-attention.
To obtain high dot-product values --- which serve as a similarity measure in the attention mechanism --- between hits from the same positron, we approached this problem by finding geometrical transformations to the hit coordinates which would efficiently align them in the embedding vector space, as will be detailed in \secref{sec:InputFeatureSPX} and \secref{sec:InputFeatureCDCH}.
The input feature vectors obtained by this geometrical transformation are then processed by the input embedding layers, which are feed-forward networks (FFNs) that transform the input feature vectors into vectors of the Transformer embedding dimension.
The output class probabilities are computed using an FFN followed by softmax functions \cite{bridle1990probabilistic}.

\subsection{Input feature from pTC}\label{sec:InputFeatureSPX}
\noindent Each pTC hit feature is represented by a 6-dimensional vector:
\begin{itemize}
  \item $z_\mathrm{counter}$: $z$ coordinate of the counter center (the center of the rectangular counter),
  \item $\phi_\mathrm{counter} - \phi_\mathrm{firstTC}$: where $\phi_\mathrm{counter}$ is the $\phi$ coordinate of the counter center and $\phi_\mathrm{firstTC}$ is that of the first pTC hit tile,
  \item $\cos(\phi_\mathrm{counter} - \phi_\mathrm{firstTC})$, $\sin(\phi_\mathrm{counter} - \phi_\mathrm{firstTC})$,
  \item $u_\mathrm{conformal}$, $v_\mathrm{conformal}$: conformal-mapped coordinates of the hit position (defined by \eqref{eq:ConformalMapping}), estimated with \SI{1}{cm} precision in \secref{sec:ConventionalReco}, and then rotated by $\phi_\mathrm{firstTC}$.
\end{itemize}

The rotation by $\phi_\mathrm{firstTC}$ takes advantage of the detector's rotational symmetry.
Although the full $\phi$ acceptance of the detector is defined as $|\phi_e| < \SI{60}{\degree}$, this rotation forces tracks to pass through $\phi = 0$ at the entry point to the pTC.
This reduces the $\phi_e$ phase space of the positron topology in the training samples, effectively increasing the number of training samples within the available phase space volume.
This rotation is also applied for hits from the CDCH, as will be discussed in \secref{sec:InputFeatureCDCH}.

\begin{figure*}[tbp]
   \centering
   \includegraphics[width=\linewidth]{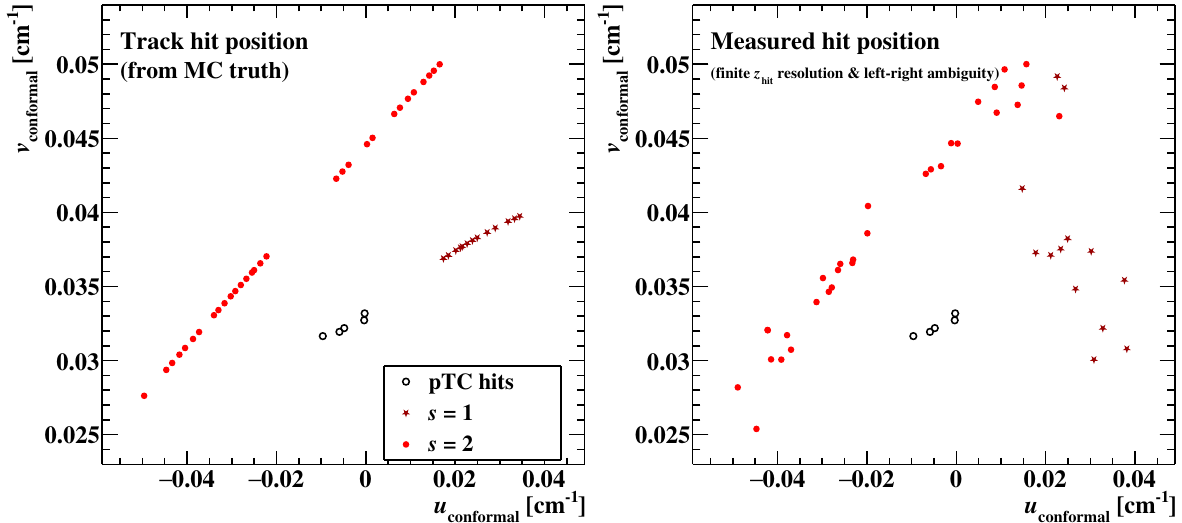}
   \caption{Distribution of hit positions for an example 1.5-turn positron track in MC samples, shown in conformal coordinates. 
            The left panel shows the true hit positions, where hits in the same turn segment are well aligned. 
            The right panel shows the reconstructed hit positions incorporating the finite detector resolution, where CDCH hit alignment deteriorates due to the limited $z_\mathrm{hit}$ resolution and the unsolved left-right ambiguity.}
   \label{fig:ConformalMapping}
\end{figure*}

The conformal mapping,
\begin{equation}\label{eq:ConformalMapping}
  (u_\mathrm{conformal}, v_\mathrm{conformal}) =
  \left(\frac{x}{x^2+y^2}, \frac{y}{x^2+y^2}\right),
\end{equation}
transforms a circle passing through the origin into a straight line in the $(u_\mathrm{conformal}, v_\mathrm{conformal})$ space.
In the COBRA magnet, positron trajectories are not perfectly circular due to the gradient magnetic field.
Nevertheless, hits within the same turn segment (i.e., the same $s$ index) are well aligned, as shown in \figref{fig:ConformalMapping}.
In particular, the pTC hits are supposed to be aligned with the $s=1$ CDCH hits.

\subsection{Input feature from CDCH}\label{sec:InputFeatureCDCH}
\noindent Each CDCH hit input feature is constructed from its coordinates and represented as a 47-dimensional vector.
Among these, 12 elements are:
\begin{itemize}
  \item $r_\mathrm{wire}$: the $r$ value of the wire at the center ($z = 0$),
  \item $d_\mathrm{drift}$, $t_\mathrm{drift}$: the estimated drift distance and time,
  \item $z_\mathrm{hit}$: $z$ from the hit reconstruction,
  \item $z_\mathrm{hit}/z_\mathrm{firstTC}$: where $z_\mathrm{firstTC}$ is the $z$ of the first pTC hit tile,
  \item $\phi_z - \phi_\mathrm{firstTC}$: where $\phi_z$ is the wire $\phi$ at $z_\mathrm{hit}$, 
  \item $\cos(\phi_z - \phi_\mathrm{firstTC})$, $\sin(\phi_z - \phi_\mathrm{firstTC})$,
  \item $u_\mathrm{conformal}$, $v_\mathrm{conformal}$: conformal coordinates of the wire at $z_\mathrm{hit}$ after rotation by $\phi_\mathrm{firstTC}$,
  \item $\frac{\D u}{\D z}$, $\frac{\D v}{\D z}$: derivatives of the conformal coordinates with respect to $z$.
\end{itemize}
The remaining 35 elements come from seven sets of 5-dimensional features, each dependent on the turn segment index described in \secref{sec:TransformerModelConcept}, $s \in \{1, 2_\mathrm{in}, 2_\mathrm{out}, 3_\mathrm{in}, 3_\mathrm{out}, 4_\mathrm{in}, 4_\mathrm{out}\}$:
\begin{itemize}
  \item $z_{\mathrm{turn};s}$, $z_{\mathrm{turn};s}/z_\mathrm{firstTC}$: where $z_{\mathrm{turn};s}$ is the typical $z$ of hits on the $s$-th turn segment,
  \item $\phi_z - \phi_\mathrm{firstTC} - \phi_{\mathrm{turn};s}$: where $\phi_{\mathrm{turn};s}$ is the typical $\phi - \phi_\mathrm{firstTC}$ of hits on the $s$-th turn segment,
  \item $\cos(\phi_z - \phi_\mathrm{firstTC} - \phi_{\mathrm{turn};s})$, $\sin(\phi_z - \phi_\mathrm{firstTC} - \phi_{\mathrm{turn};s})$.
\end{itemize}

The conformal coordinate calculable at the hit reconstruction level has a limited resolution because of the limited $z_\mathrm{hit}$ resolution and the left-right ambiguity.
The $z_\mathrm{hit}$ resolution of \SI{7.5}{cm} (see \secref{sec:SpectrometerHardware}) on the stereo-configured wires results in a few centimeter resolution in the $(x,y)$ estimate.
In addition, due to the unsolved left-right ambiguity at the hit level, the conformal coordinate is naively calculated from the wire position at $z_\mathrm{hit}$.
Without incorporating the drift distance, we have an additional $(x,y)$ uncertainty of \SIrange{6}{8}{mm}, which is the typical drift cell size.
In \figref{fig:ConformalMapping}, we see that the alignment quality on the conformal coordinate is limited at the hit level, which is particularly significant for $s=1$ hits because of the worse $z_\mathrm{hit}$ resolution for hits closer to the wire ends.
As a possible compensation measure, we included $\frac{\D u}{\D z}$ and $\frac{\D v}{\D z}$ in the feature vector.

\begin{figure*}[tbp]
  \centering
  \includegraphics[width=\linewidth]{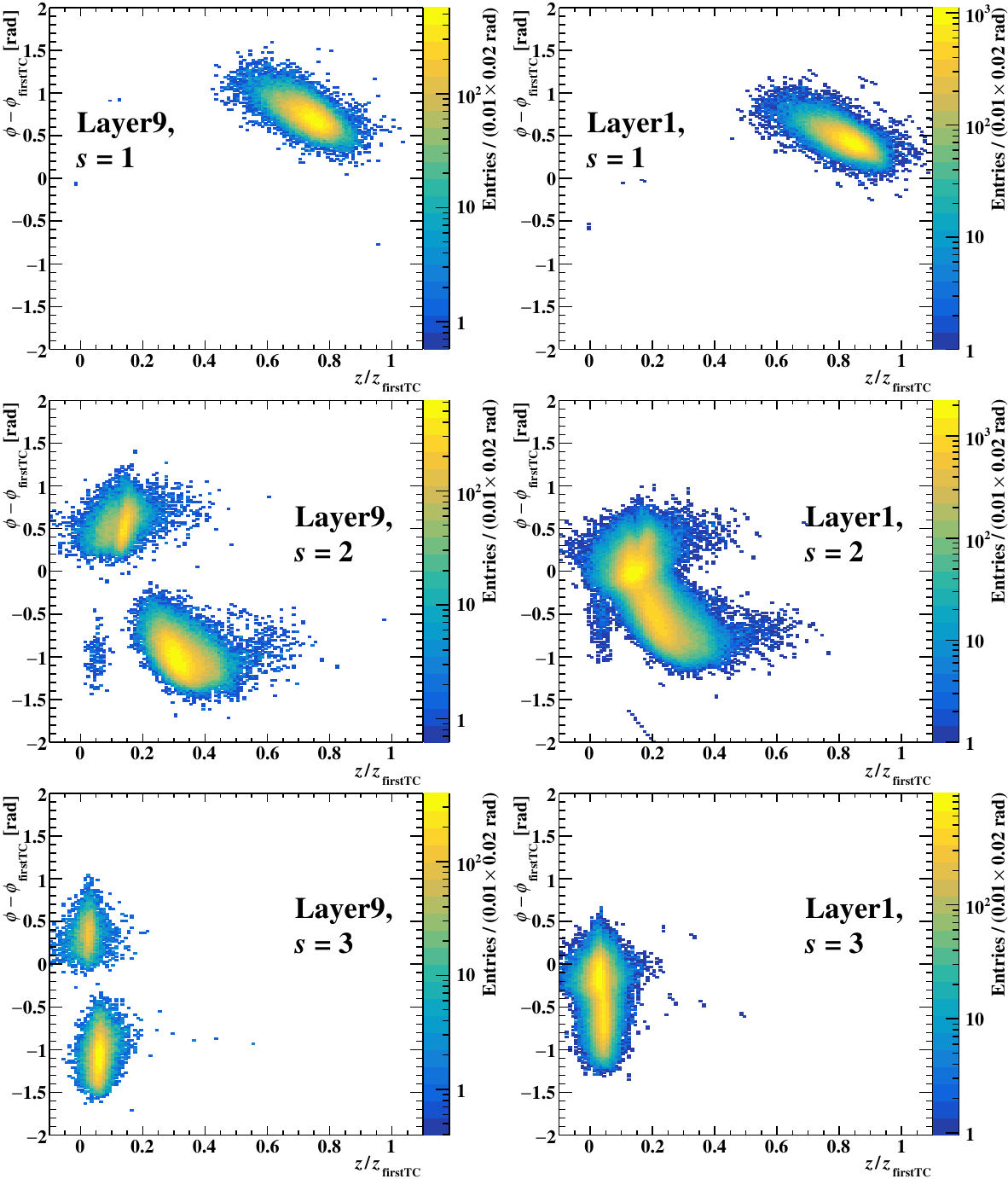}
  \caption{Distribution of $\phi$ vs.\ $z$ in data samples, which was used for $\phi_{\mathrm{turn};s}$ and $z_{\mathrm{turn};s}$ calibration, shown for different combinations of layer and $s$ index. Track-fitted $\phi$ and $z$ coordinates of each hit are used in these plots. The left (right) three plots correspond to the innermost (outermost) layer.}
  \label{fig:PhiZPattern}
\end{figure*}

The $s$-dependent $z_{\mathrm{turn};s}$ and $\phi_{\mathrm{turn};s}$ patterns were introduced to leverage dot-product attention between distant hits belonging to different turn segments.
These patterns were calibrated using Michel positron tracks, as shown in \figref{fig:PhiZPattern}.
The use of $z_{\mathrm{turn};s}/z_\mathrm{firstTC}$ is motivated by its narrower track-by-track variation compared to the simple $z_{\mathrm{turn};s}$ distribution, as well as its symmetry between downstream (positive $z$) and upstream (negative $z$).
The dependence of this pattern on the drift-cell layer, as seen in \figref{fig:PhiZPattern}, was also taken into account when constructing the feature vector.
\begin{figure}[tbp]
  \centering
  \includegraphics[width=\linewidth]{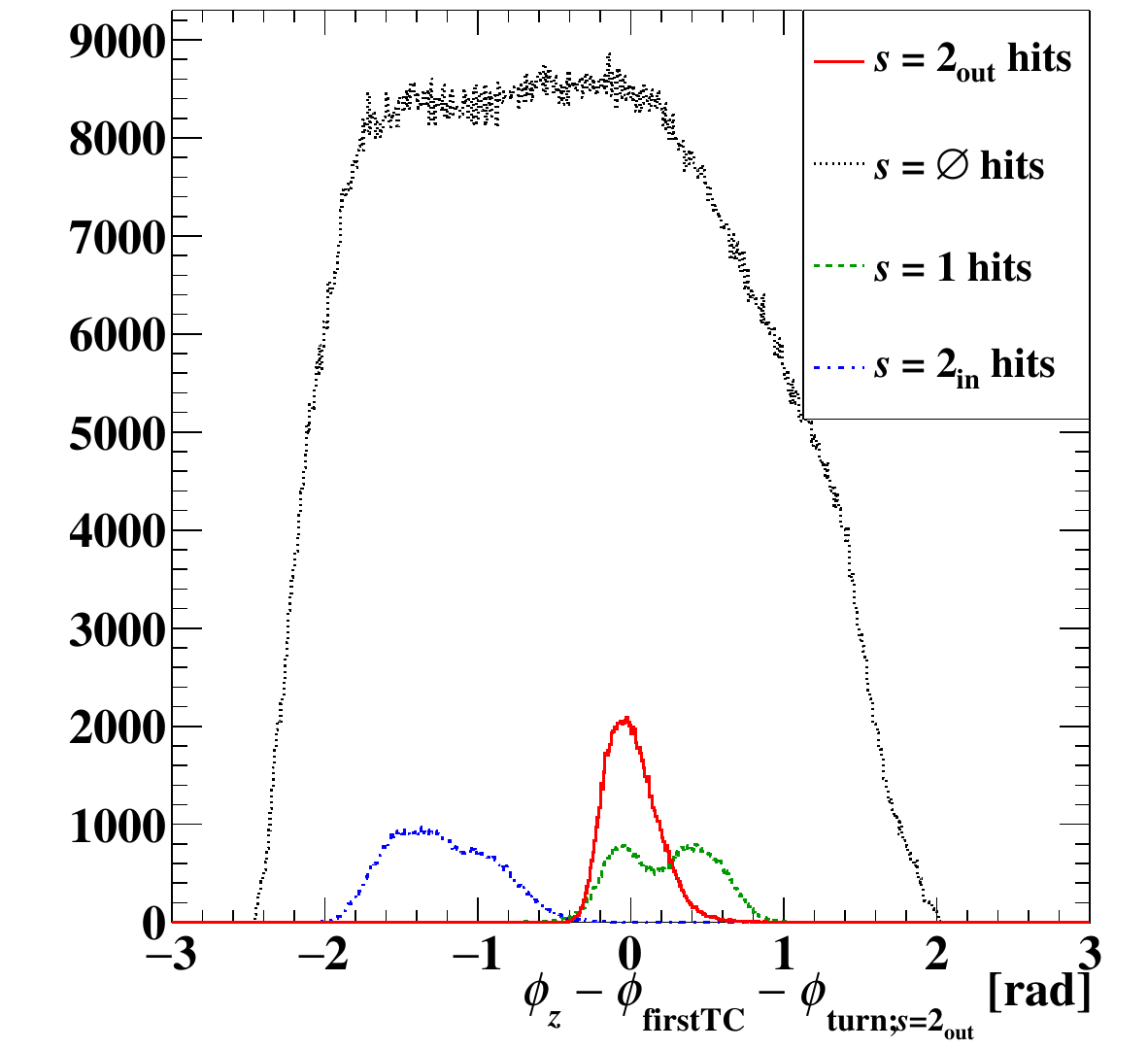}
  \caption{Distribution of $\phi_z - \phi_\mathrm{firstTC} - \phi_{\mathrm{turn};s=2_\mathrm{out}}$ for $s=\varnothing, 1, 2_\mathrm{in},$ and $2_\mathrm{out}$ hits. 
           The narrower distribution for $s=2_\mathrm{out}$ indicates that the correction of $\phi_{\mathrm{turn};s=2_\mathrm{out}}$ can efficiently extract the feature of $s=2_\mathrm{out}$ hits out of the others.
           The total number of entries for each $s$ label is normalized to the hit rate at \SI{5e7}{\mu/s} rate, where $s=\varnothing$ hits (i.e. pileup) accounts for \SI{90}{\percent} of hits within a \SI{300}{ns} window.
  }
  \label{fig:PhiCorrectedDistribution}
\end{figure}
The $\phi_{\mathrm{turn};s=2_\mathrm{out}}$-corrected distribution is compared between $s=\varnothing, 1, 2_\mathrm{in}, 2_\mathrm{out}$ in \figref{fig:PhiCorrectedDistribution}, where hits from $2_\mathrm{out}$ shows narrower distribution than the other classes of hits.

\subsection{Training scheme}
\noindent The training dataset, consisting of one million samples, was primarily based on MC samples of both \SI{52.8}{MeV} and Michel positrons.
As the model inference is given for each pTC cluster (See \secref{sec:TransformerModelConcept}), one training sample consists of a set of hits from the associated positron trajectory (CDCH hits annotated as $s\neq\varnothing$ and pTC hits in the cluster of interest) and a set of pileup CDCH hits ($s=\varnothing$).
Here, the pileup hits are created by different close-in-timing positrons that leave no pTC hit (if outside the pTC acceptance) or different pTC clusters.
With the MC samples, hits in the training data can be correctly annotated by the MC truths.
However, the detector alignment assumed in the MC production can be different from that in the data within the calibration precision of a few hundred \SI{}{\micro\meter}.
With concerns that the model may overfit to the specific alignment assumed in the MC samples, we included Michel positron data samples collected at low muon stopping rate (less than \SI{e6}{\mu/s}).
In these data samples, the CDCH hit labels were annotated by the result of the conventional track reconstruction described in \secref{sec:ConventionalReco}.
According to our simulations at such a low rate, the conventional reconstruction can correctly label the $s$-indices for more than \SI{95}{\percent} of CDCH hits of reconstructed tracks.
Though imperfect, we judged that the quality of this annotation with data samples is enough as supplementary training samples.
To emulate a high pileup environment, these data-driven annotations at low intensity were then event-mixed with CDCH hits in data samples collected for background studies (triggered at random time), which were labelled as $s=\varnothing$.
MC samples (complementary data samples) accounted for \SI{90}{\percent} (the remaining \SI{10}{\percent}) of the entire training dataset.
We prepared the training samples at four different pileup rates, which correspond to the muon stopping rate of below \SI{e6}{\mu/s}, \SI{3e7}{\mu/s}, \SI{4e7}{\mu/s}, and \SI{5e7}{\mu/s}.

The model, which outputs multiple class labels, was trained using a weighted cross-entropy loss function, where class-specific weights depend on the target label $s$.
These weights were introduced to compensate for the imbalance in sample counts across different class labels, as $s = \varnothing$ (pileup or noise) dominates.
We split the training process into the following two steps.
In the first step, the model was trained only with samples at the low pileup rate, namely the one corresponding to the muon stopping rate below \SI{e6}{\mu/s}.
In the following second step, all the samples, including pileup hits at various muon stopping rates, \SIrange{3}{5e7}{\mu/s}, were used to fully tune the model parameters.
With the split training procedure, we observed faster convergence of the loss value than a single step training using the high-rate samples from the beginning.
In our understanding, the first step allowed the model to efficiently learn the correlation between $s\neq\varnothing$ hits, which simplifies the second step as learning how to discriminate $s=\varnothing$ from $s\neq\varnothing$ hits.

The model hyperparameters were decided mainly considering the available RAM resource in the offline data processing running on CPU machines (mentioned later in \secref{sec:SoftwareImplementation}), where we aimed to contain the whole tracking algorithm within \SI{2}{GB}.
The trained model in this study contains 35 million parameters (i.e., trainable weights and biases), with the total model size being \SI{140}{MB}.
We compared this choice with different choices with increased model complexity (the number of attention heads, the number of layers, and the dimension of hidden layers), observing no further improvement in the loss value.
On the other hand, the model size of \SI{140}{MB} is not a major part of the allowed \SI{2}{GB} space (Kalman filter algorithms in the track fitting dominate after all), not demanding intensive studies for the size reduction.
We therefore decided to deploy the model with a reasonable model size, which also leaves little room for further improvement in the hit filtering performance.

\subsection{Efficiency and purity}\label{sec:EfficiencyPurity}

\begin{figure}[tbp]
  \centering
  \includegraphics[width=\linewidth]{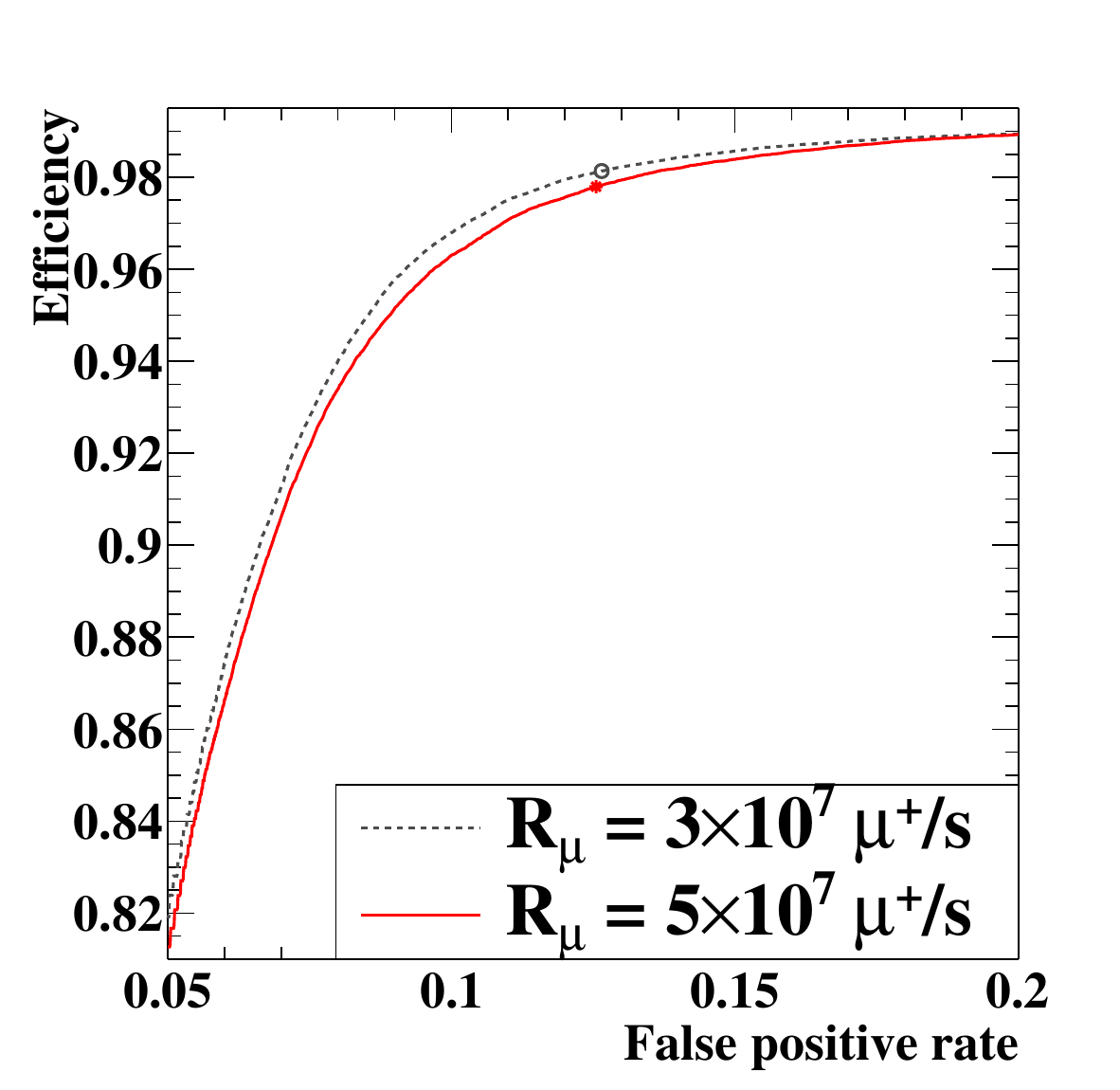}
  \caption{Correlation between signal-hit efficiency and false positive rate for pileup in the validation dataset at \SI{5e7}{\mu/s} (red solid line) and \SI{3e7}{\mu/s} (black dashed line). The marker on each line indicates the values at the adopted ML output threshold of $f_\mathrm{th}=0.95$.}
  \label{fig:ROCCurve}
\end{figure}

\begin{figure}[tbp]
  \centering
  \includegraphics[width=\linewidth]{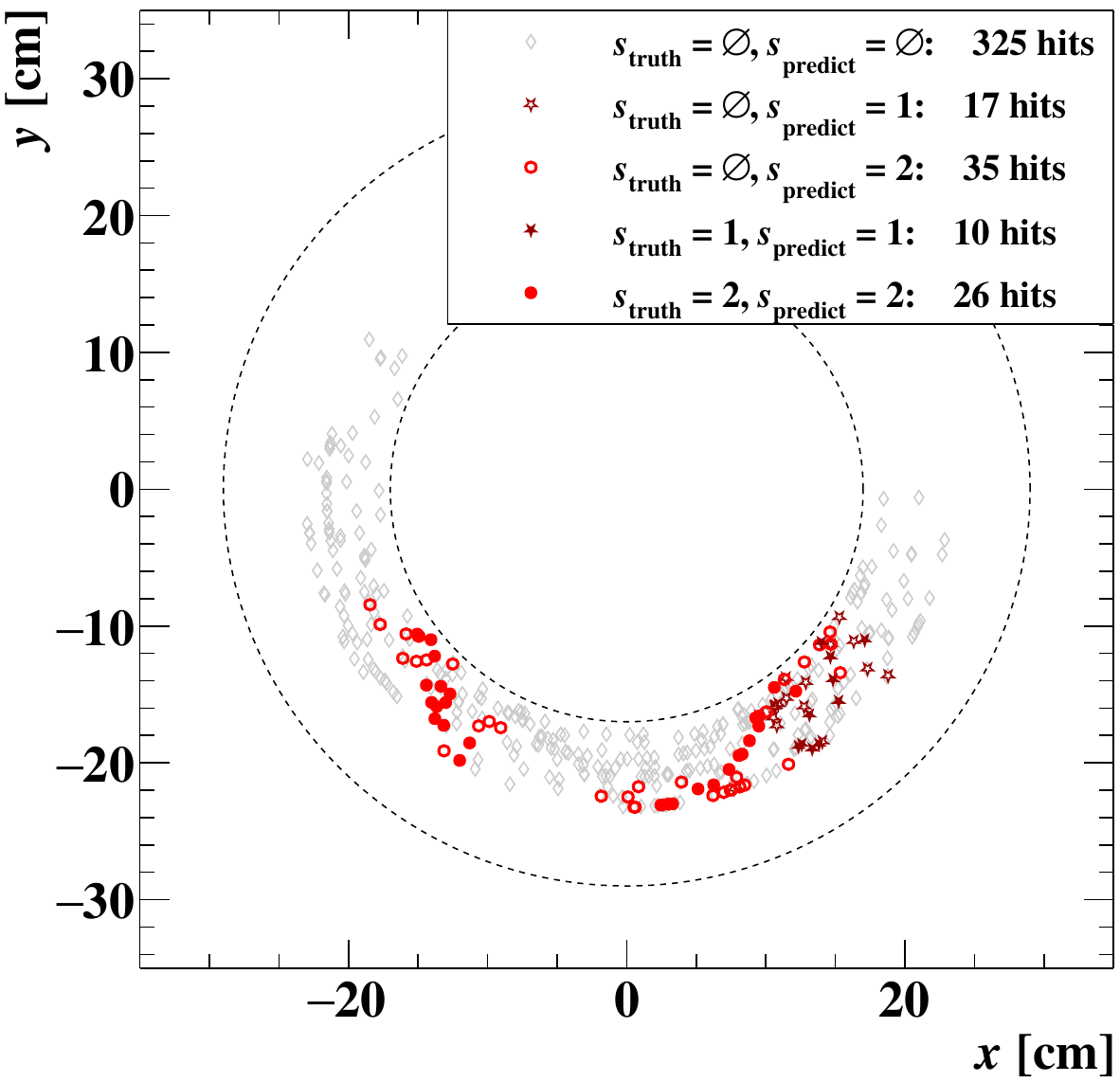}
  \caption{Model outputs for an example 1.5-turn track in the validation dataset, compared with the MC truth. 
           The $s_\mathrm{truth}$ and $s_\mathrm{predict}$ in the legend indicate the label with the MC truth and that predicted by the model, respectively.
           The $s_\mathrm{truth}=\varnothing$ hits are pileup hits simulated with the rate of \SI{5e7}{\mu/s}.
           In this event, all signal hits (36 hits in total) are correctly predicted, while 52 out of 377 pileup hits become false positives. 
  }
  \label{fig:TransformerDisplay}
\end{figure}

\noindent The signal-hit efficiency and the false positive rate for pileup were studied at various thresholds of the ML output.
Their values in the validation dataset at muon stopping rates of \SI{3e7}{\mu/s} and \SI{5e7}{\mu/s} are shown in \figref{fig:ROCCurve}.
The selected threshold ($f_\mathrm{th}$ in \eqref{eq:HitFilteringCriteria}) was obtained by scanning the tracking efficiency at different $f_\mathrm{th}$ values, with the scan interval of $f_\mathrm{th}$ being 0.025.
The markers in \figref{fig:ROCCurve} indicate the performance at the adopted threshold: \SI{98}{\percent} efficiency and \SI{13}{\percent} false positive rate at $f_\mathrm{th}=0.95$.
With the marginal dependences of purity and efficiency on the muon beam rate, the pileup hit occupancy is proportional to the muon beam rate.
Here, as presented in \figref{fig:PhiCorrectedDistribution}, the pileup hits count up nine times more than signal hits at \SI{5e7}{\mu/s}.
When the filtering is applied at this rate, \SI{50}{\percent} of the filtered hits are signal hits from trajectories of interest, which thereby benefits the subsequent track candidate construction and fitting.
Examples of ML-based hit selection at the adopted threshold (for \SI{5e7}{\mu/s}) are shown in \figref{fig:TransformerDisplay}.

\subsection{Software implementation}\label{sec:SoftwareImplementation}
\noindent The Transformer model was trained using \emph{PyTorch}, while the offline reconstruction for the MEG~II experiment is implemented in C++.
The trained model was integrated into the C++ code with the aid of the ONNX framework \cite{onnx2017}, and model inference during reconstruction runs on CPU processors.

\section{Tracking performance improvement}\label{sec:Performance}
\noindent As discussed in \secref{sec:TransformerModelConcept}, we updated the track reconstruction, with the ML-based hit filtering applied before running the track finding and fitting algorithms.
In the following subsections, we present the tracking efficiency, resolution, and CPU efficiency obtained with the updated method and compare them with those without hit filtering.

\subsection{Efficiency}
\begin{figure}[tbp]
  \centering
  \includegraphics[width=\linewidth]{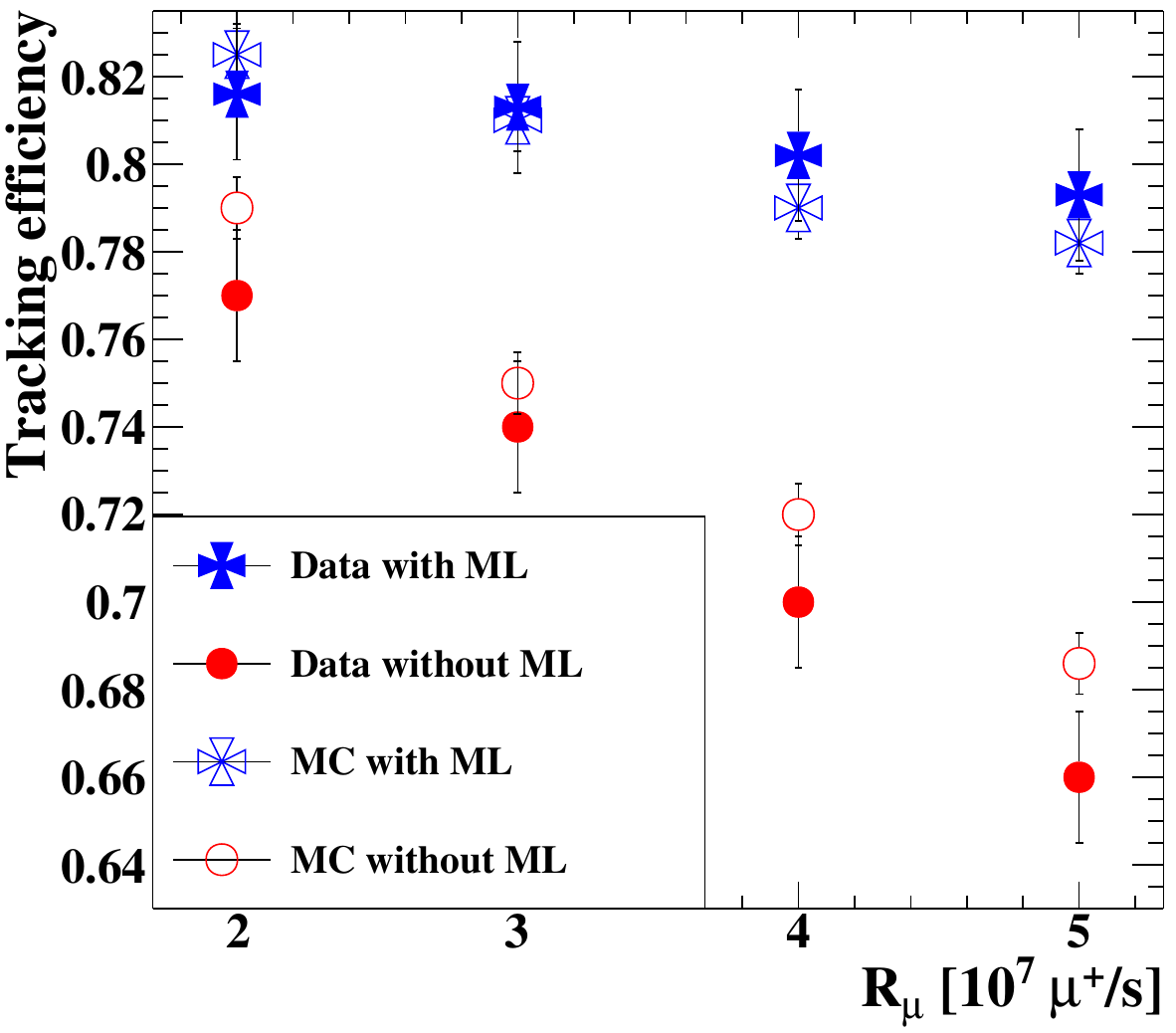}
  \caption{Comparison of tracking efficiency with and without the use of ML in the reconstruction for MC and data samples. 
           The efficiency without ML corresponds to the result presented in \citeref{MEGIIDetectorPaper2023}.
           The error bars for the data are dominated by the uncertainty in the beam rate estimation discussed in \citeref{MEGIIDetectorPaper2023}, where the errors at different rates are fully correlated to each other.
           The error bars for the MC samples are from the statistical uncertainty with the produced number of MC samples, and the errors are independent of each other.
  }
  \label{fig:CDCHEfficiencyML}
\end{figure}

\noindent The impact on tracking efficiency was evaluated using the positron counting method described in \secref{sec:ConventionalPerformance}.
\figref{fig:CDCHEfficiencyML} compares the efficiencies of the conventional and proposed methods at different muon stopping rates, ranging between \SIrange{2}{5e7}{\mu/s}.
We observed a larger improvement in efficiency at higher muon stopping rates.
In particular, an efficiency gain of \SI{15}{\percent} was achieved at \SI{5e7}{\mu/s}.
These results obtained from data were cross-checked with MC samples, showing agreement within \SI{1}{\percent}.

As discussed in \secref{sec:EfficiencyPurity}, the hit filtering performs with the hit level signal efficiency of \SI{98}{\percent} while discarding \SI{87}{\percent} of pileup hits.
The impact of the inefficient \SI{2}{\percent} hits on the tracking efficiency was studied by reconstructing tracks in MC samples at \SI{5e7}{\mu/s} with the following two configurations:
\begin{enumerate}
 \item Hit selection by ML without any modification (\SI{98}{\percent} efficiency and \SI{13}{\percent} false positive).
 \item After selection by ML, mis-identified signal hits are recovered by means of MC truth (\SI{100}{\percent} efficiency and \SI{13}{\percent} false positive).
\end{enumerate}
As a result, we observed an approximately \SI{1}{\percent} higher tracking efficiency with the recovery of \SI{2}{\percent} signal hits. 
Nevertheless, this difference is not as large as the improvement presented in \figref{fig:CDCHEfficiencyML}, which is the difference between the first configuration and no hit filtering (i.e. \SI{100}{\percent} efficiency with \SI{100}{\percent} false positive).
Therefore, the tracking efficiency improvement can be attributed to the reduced pileup hits, while the \SI{2}{\percent} signal hit inefficiency potentially leaves room for an additional \SI{1}{\percent} improvement.

In addition to the direct comparison of the tracking efficiency, we studied the number of CDCH hits on fitted tracks that match well the MC truth (i.e., hits that are actually on the trajectory of interest, not that of a pileup).
\begin{figure}[tbp]
  \centering
  \includegraphics[width=\linewidth]{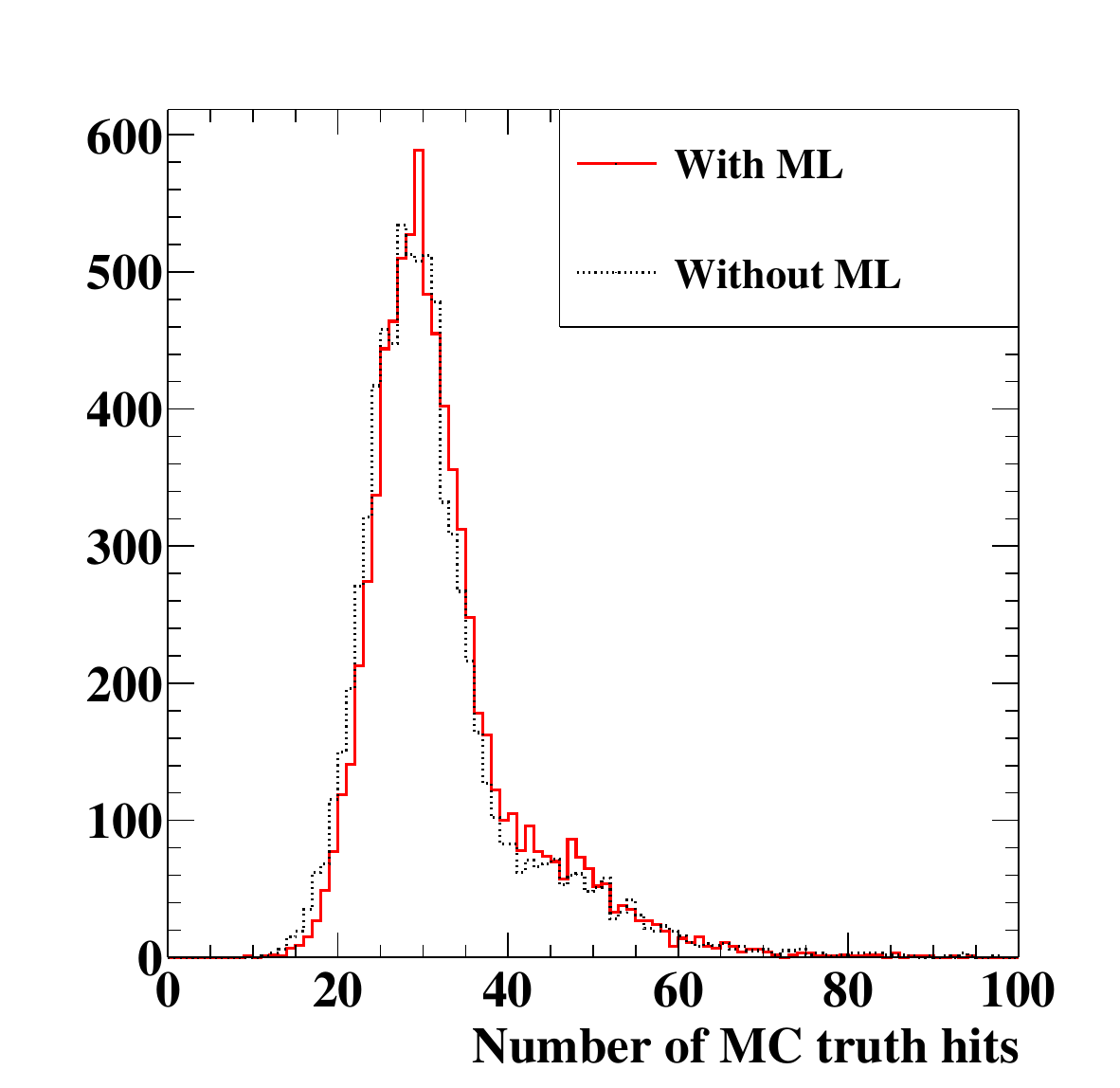}
  \caption{The number of hits on the fitted tracks actually associated with MC truth.
           The distribution is compared between the reconstruction with (red solid line) and without (black dotted line) ML-based hit filtering.}
  \label{fig:NTrueHits}
\end{figure}
In \figref{fig:NTrueHits}, we see an improvement thanks to the ML-based hit filtering. 
This again indicates that the tracking algorithm can efficiently identify hits from trajectories of interest with the reduced number of pileup hits, which the authors understand contributes to the tracking efficiency improvement.

\subsection{Resolution}
\begin{figure}[tbp]
  \centering
  \includegraphics[width=\linewidth]{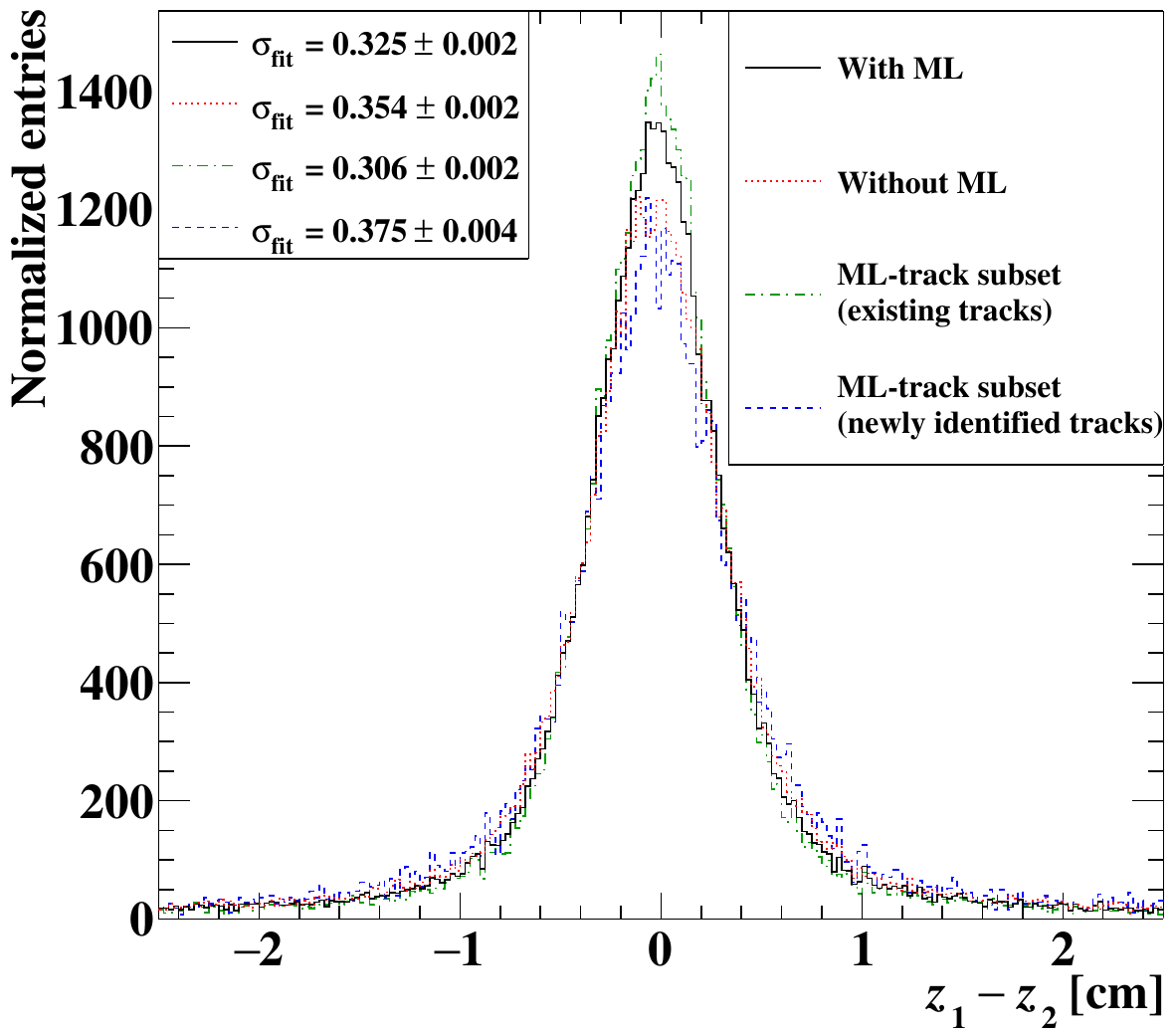}
\caption{
      Comparison of $z$-resolution obtained in the double-turn analysis between the conventional tracking algorithm and the ML-based approach, including detailed comparisons between subsets of ML-reconstructed tracks: those also found by the conventional algorithm and those newly identified.
      The graphs show the $z_1 - z_2$ distribution, where $z_1$ and $z_2$ correspond to independent fits of the first and second turns of split double-turn tracks.
      The histograms were scaled so that all have equal area, providing a clearer visual comparison of resolution.
      The $\sigma_\mathrm{fit}$ shown in the legend is the standard deviation of each distribution obtained by a Gaussian fitting.
      Results indicate an overall \SI{5}{\percent} improvement in resolution with ML tracking, and an even greater \SI{10}{\percent} improvement for tracks already reconstructed by the conventional algorithm.
}
  \label{fig:DoubleTurnZImprovementML}
\end{figure}

\noindent \figref{fig:DoubleTurnZImprovementML} illustrates the $z$-resolution improvement observed using the double-turn analysis method introduced in \secref{sec:ConventionalPerformance}.
The plot shows the distribution of differences between independently fitted track $z$ values for the first and later turns of split double-turn tracks.
The overall $z$-resolution, estimated by fitting a Gaussian to the distribution and correcting the differences between the original tracks and the split tracks (see Sec.~11.1 of \citeref{MEGIICDCHPaper2023} for the full procedure), improved by \SI{5}{\percent} with the Transformer-based model.
Similar analyses of other position ($y$) and angle variables ($\phi$ and $\theta$) consistently showed improvements of approximately \SI{5}{\percent} overall.

\figref{fig:DoubleTurnZImprovementML} also compares the $z$-resolution for subsets of ML-reconstructed tracks: those also found by the conventional algorithm and those newly identified (i.e., those contributing to the efficiency improvement shown in \figref{fig:CDCHEfficiencyML}).
ML-reconstructed tracks that were also found by the conventional algorithm exhibit even better resolution, \SI{10}{\percent} higher than the average resolution of the conventional algorithm.
Conversely, tracks newly identified by the ML approach show \SI{10}{\percent} worse resolution than the conventional average.
This indicates that ML-based tracking improves efficiency primarily by recovering low-quality tracks that were difficult for the conventional algorithm to reconstruct.
These improvements in efficiency and resolution are understood to result from the purity enhancement presented in \figref{fig:ROCCurve}.

\begin{figure}[tbp]
  \centering
  \includegraphics[width=\linewidth]{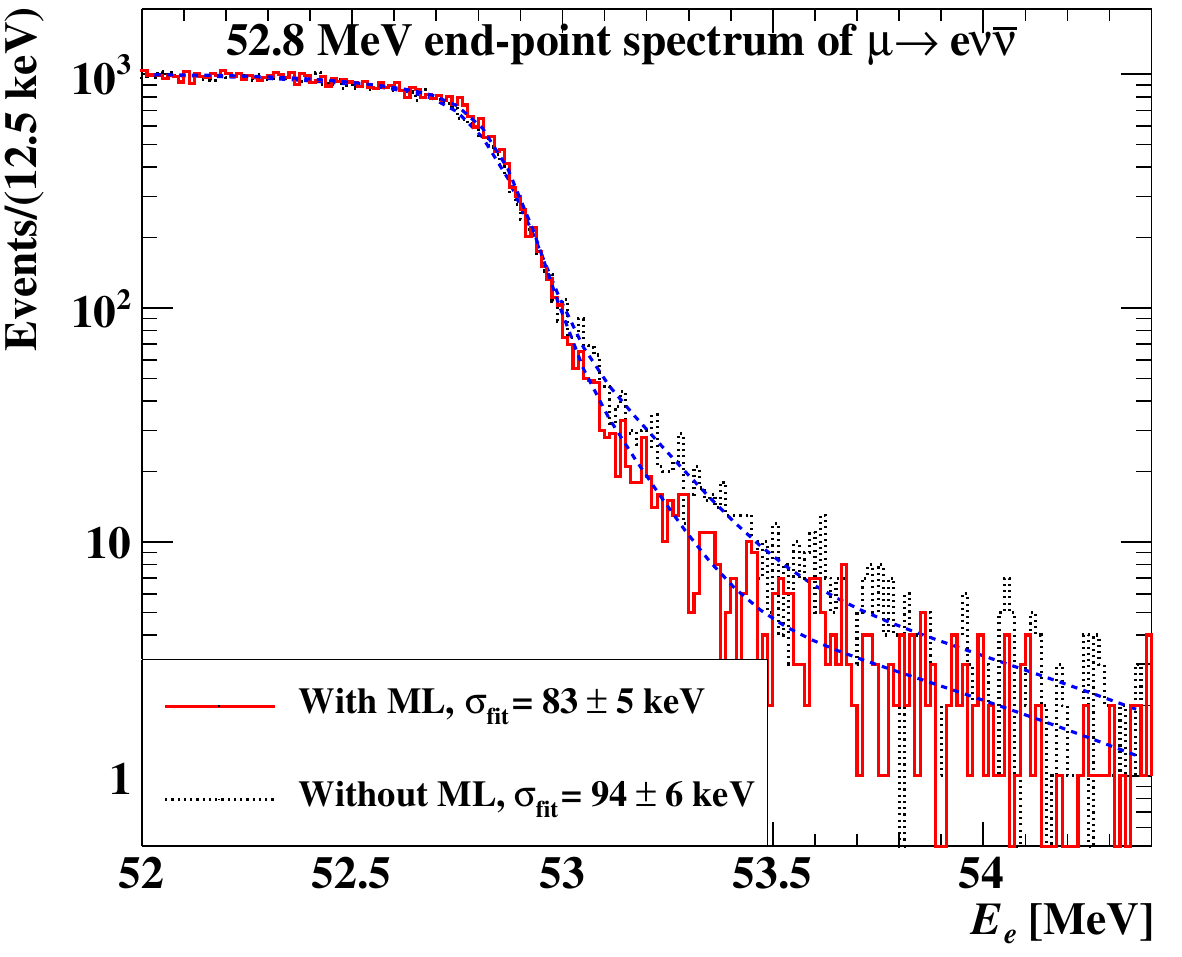}
  \caption{
    Comparison of the \SI{52.8}{MeV} endpoint spectrum for the same set of tracks reconstructed by both the ML-based and conventional algorithms.
    The number of tracks above \SI{52.8}{MeV}, which is kinematically forbidden, is smaller with ML reconstruction, indicating an improvement in $E_e$ resolution.
    The blue dashed lines show the fitted line, obtained by convolving the theoretical spectrum with the finite detector resolution.
    The $\sigma_\mathrm{fit}$ in the plot indicates the fitted value of the detector resolution.
  }
  \label{fig:MichelEdgeImprovementMLSubset}
\end{figure}

The improvement in momentum resolution can be observed in the \SI{52.8}{MeV} endpoint spectrum of Michel positrons.
\figref{fig:MichelEdgeImprovementMLSubset} compares the reconstructed positron energy spectra obtained with the ML-based and conventional methods.
As shown, the number of tracks exceeding \SI{52.8}{MeV} is reduced with ML reconstruction.
Since the theoretical endpoint lies strictly at \SI{52.8}{MeV}, tracks with $E_e > \SI{52.8}{MeV}$ originate from the finite tracking resolution.
Therefore, this demonstrates the improvement in momentum resolution.
The observed end-point spectra were fitted with the theoretical spectrum convolved with a sum of Gaussian modeling of the detector resolution (introduced in \secref{sec:ConventionalPerformance} and detailed in Sec.~11.1 of \citeref{MEGIICDCHPaper2023}). 
The fitted values of the detector resolution indicate an approximately \SI{10}{\percent} improvement with the ML-based tracking approach.

\subsection{CPU time}
\noindent \tabref{tab:CPUTime} shows the computing time necessary to process one event.
With the hit filtering by the Transformer-based model, the total computing time for the entire positron reconstruction sequence was reduced to \SIrange{70}{80}{\percent} of that required by the conventional approach.
In particular, the cost of track seeding and candidate construction --- steps that are especially sensitive to pileup occupancy --- was halved thanks to the improved hit purity.
Nevertheless, the track finding and subsequent fitting algorithms still dominate the total computing time, while the model's inference accounts for \SI{5}{\percent}.
\begin{table}[tbp]
  \centering
  \caption{Time to process one event for both reconstruction (for pileup rate of \SI{5e7}{\mu/s}) and simulation with AMD EPYC 7713, and its breakdown into sub-tasks. 
           Note that when simulating one $\meg$ event, one $\meg$ sample is event-mixed with O(100) beam muon simulation samples.}
  \label{tab:CPUTime}
  \vspace{0.5em}
  \begin{tabular}{lcc}
  \noalign{\smallskip}
  \hline
  Task & Without ML & With ML \\
  \hline\hline\noalign{\smallskip}
  \multicolumn{3}{c}{Positron track reconstruction time (\SI{5e7}{\mu/s})} \\
  \hline
  Waveform analysis & \multicolumn{2}{c}{3 s} \\
  ML inference & --- & 1 s \\
  Seed \& candidate construction & 15.5 s & 7.5 s \\
  Track fitting \& refinement & 4 s & 6 s \\
  \hline
  Total & 22.5 s & 17.5 s\\
  \hline\hline\noalign{\smallskip}
  \multicolumn{3}{c}{Simulation time} \\
  \hline
  Single $\meg$ generation    & \multicolumn{2}{c}{17 s} \\
  Single beam muon generation & \multicolumn{2}{c}{3.5 s} \\
  Event mixing & \multicolumn{2}{c}{6 s} \\
  \hline
  \end{tabular}
\end{table}
In \tabref{tab:CPUTime}, the increase in the computing time for the track fitting arises from the additional post-fit track refinement using the ML-prediction (mentioned in \secref{sec:TransformerModelConcept}), as well as the increase in the number of tracks to be fitted.

\section{Discussion}\label{sec:Discussion}
\noindent The Transformer-based model improved both the efficiency and resolution of positron measurements, as presented in the previous section.
Both improvements are understood to result from enhanced hit purity during track finding and track fitting.
In track finding, improved purity facilitates the construction of correct seeds, leading to higher efficiency.
In track fitting, residual impurities are understood to remain even after applying the DAF, introducing additional errors in determining track kinematics.
Thus, removing such impure hits also contributes to improved tracking resolution.

With these performance improvements, the MEG~II experiment has decided to apply this algorithm to the already collected dataset, including reprocessing the data previously analyzed in \citeref{MEGIIResult2025}.
In addition, the muon stopping rate has been increased to \SI{5e7}{\mu/s} since the start of the 2025 DAQ as a result of re-optimization.
Together, these two updates are expected to improve the experimental sensitivity by approximately \SI{10}{\percent}.

Given the highly encouraging results presented in this paper, it would be valuable to further extend the model for application to a broader range of tasks.
Currently, the most promising extension is to develop an end-to-end model capable of directly estimating track kinematics while resolving hit-level ambiguities.
Such a model could improve track candidate construction, thereby reducing CPU time and increasing tracking efficiency.
Moreover, with better-constructed track candidates, more accurate initial values for DAF-based track fitting could be obtained, potentially leading to further improvements in tracking resolution.

\section{Conclusion}\label{sec:Conclusion}
\noindent In the MEG~II experiment searching for $\meg$, efficient and precise positron measurements are essential to achieve high sensitivity.
In this context, this paper presented a successful application of a Transformer-based model to enhance positron tracking performance.
The tracking algorithm must fully reconstruct multiple turns under a high pileup occupancy of \SIrange{35}{50}{\percent} in the CDCH.
To address this challenge, the model was designed as a classifier to filter out pileup hits.
At the selected threshold on the Transformer outputs, the model achieved \SI{98}{\percent} efficiency for signal hits while discarding \SI{87}{\percent} of pileup hits, providing a purity-enhanced set of hits for the track seeding algorithm.
This contributed to improvements in positron tracking efficiency and resolution by \SI{15}{\percent} and \SI{5}{\percent}, respectively, compared to the results reported in \citeref{MEGIICDCHPaper2023}. 
Given these performance gains, the experiment has decided to increase the muon stopping rate and reprocess the already collected dataset.
As a result, the sensitivity of the $\meg$ branching ratio measurement is expected to improve by approximately \SI{10}{\percent}, representing a significant scientific gain.
In the next MEG~II publication on the search for $\meg$, positrons will be reconstructed using the method described in this paper.

\section*{Acknowledgement}
\noindent We gratefully acknowledge the technical support and cooperation provided by PSI as the host institute.
This work particularly benefited from access to GPU resources on the \emph{gMerlin} cluster.
We also thank the MEG~II collaboration members for ensuring stable detector operation and for collecting the dataset used in this study.
This work was supported by JSPS KAKENHI Grant Numbers 21H04991 and 22K21350 (Japan); Leverhulme Trust LIP-2021-014 (UK).

\bibliography{mybibfile}

\end{document}